\documentclass[a4paper,fleqn,usenatbib]{mnras}

\usepackage[T1]{fontenc}
\usepackage{ae,aecompl}

\usepackage{graphicx}	
\usepackage{amsmath}	
\usepackage{amssymb}	

\title[Pop~III Stars and the 21-cm Signal]{The Effects of Population III Radiation 
Backgrounds on the Cosmological 21-cm Signal}

\author[R. H. Mebane et al.]{
Richard H. Mebane,$^{1}$\thanks{E-mail: rmebane@astro.ucla.edu}
Jordan Mirocha,$^{2, 3}$
Steven R. Furlanetto$^{1}$
\\
$^{1}$Department of Physics \& Astronomy, University of California, Los Angeles, Los Angeles, CA 90095, USA
\\
$^{2}$Department of Physics and McGill Space Institute, McGill University, Montreal QC H3A 2T8, Canada
\\
$^{3}$CITA National Fellow
}

\date{Accepted XXX. Received YYY; in original form ZZZ}

\pubyear{2019}

\begin{document}
\label{firstpage}
\pagerange{\pageref{firstpage}--\pageref{lastpage}}
\maketitle

\begin{abstract}
We investigate the effects of Population III (Pop~III) stars and their remnants on the cosmological 21-cm global signal. By combining a semi-analytic model of Pop~III star formation with a
global 21-cm simulation code, we investigate how X-ray and radio emission from accreting Pop~III black holes 
may affect both the timing and depth of the 21-cm absorption feature that follows the initial onset of star formation 
during the Cosmic Dawn. 
We compare our results to the findings of the EDGES experiment, which has reported the first detection of a cosmic 21-cm signal. In 
general, we find that our fiducial Pop~III models, which have peak star formation rate densities of 
$\sim 10^{-4}$ M$_\odot$ yr$^{-1}$ Mpc$^{-3}$ between $z \sim 10$ and $z \sim 15$, are able to match the timing of the EDGES signal quite well,
in contrast to models that ignore Pop~III stars. To match the unexpectedly large depth of the EDGES signal
without recourse to exotic physics, we vary the parameters 
of 
emission from accreting black holes (formed as Pop III remnants) including the intrinsic strength of X-ray and radio emission as well as the 
local column density of neutral gas. 
We find that models with strong radio emission and relatively weak X-ray emission
can self-consistently match the EDGES signal, though this solution requires 
fine-tuning. We are only able to produce signals with sharp features similar to the EDGES signal if the Pop~III IMF 
is peaked narrowly around $140 \, M_\odot$.
\end{abstract}



\section{Introduction}

The first generation of stars was formed from pristine gas with zero metallicity, and their properties were likely very different from the stars that form today. These early stars, known as 
Population~III (Pop~III) stars, are thought to have been very massive and luminous due to the decreased efficiency of molecular hydrogen cooling in metal-free minihalos \citep[][]{bromm_1999, abel_2002, bromm_2013}. Since their birth 
halos were likely very small with relatively low binding energies, feedback from this form of star formation likely played a crucial role in limiting them to only form in very small clusters 
or even in isolation \citep[][]{machacek_2001, wise_abel_2007, oshea_norman_2008, shapiro_2004, visbal_2017}. Despite this, however, they must have played a vital role in the evolution of early galaxies as they were the first sources of metals required for more traditional star 
formation.

As there have been no observations of a Pop~III star forming halo, most studies have been in the form of analytical models (e.g., \citealt{mckee_2008}; \citealt{kulkarni_2013}), numerical simulations (e.g., \citealt{machacek_2001}; \citealt{wise_abel_2007}; \citealt{oshea_norman_2008}; \citealt{xu_2016}; \citealt{stacy_2012}; \citealt{hirano_2015}; \citealt{maio_2010}; \citealt{sarmento_2018}), or semi-analytic models 
(e.g., \citealt{trenti_2009}; \citealt{crosby_2013}; \citealt{jaacks_2017}; \citealt{visbal_2017}; \citealt{mebane_2018}). The parameters of this mode of star formation are largely unconstrained, and many models predict that it will be nearly impossible to directly observe a Pop~III star, even with 
the next generation of space telescopes such as JWST. Because of this, we may have to look to indirect 
observations such as their supernovae or their effect on the cosmological 21-cm background.

The prospect of indirectly detecting Pop~III stars through the cosmic 21-cm signal is strengthened by the recent claimed first detection of 
such a signal by the EDGES experiment last year \citep[][]{edges}. 
Their claimed detection was somewhat anomalous when compared to current theoretical predictions of the 21-cm signal. In particular, the trough was deeper than expected, and the timing of the signal was inconsistent with empirically-calibrated models of high-$z$ galaxies extrapolated down to the atomic cooling 
threshold \citep{mirocha_2019}. There has been much discussion of this signal, and potential explanations include, for example, exotic physics, a radio background in excess to the CMB at these redshifts, and new modes of star formation such as metal-free Pop~III stars \citep[see, for example][]{barkana_2018, slatyer_2018, hirano_2018, munoz_2018, berlin_2018, kovetz_2018, cheung_2018, moroi_2018, chianese_2019, falkowski_2018, lawson_2019, jia_2019, costa_2018}. 
One intriguing explanation is that there is a previously unexplained radio background present at high redshift that dominates over the CMB \citep{feng_2018}, which requires that only 10\% of the excess radio background reported by the ARCADE 2 experiment in \citet{arcade2} is produced at very high redshifts \citep[e.g.,][]{aew_2018, fialkov_2019}.

In this paper we investigate the effect of Pop~III stars on the global 21-cm signal by combining the Pop~III semi-analytic model described in \citet{mebane_2018} with the 21-cm global signal 
simulation code \textsc{ares} \citep[][]{ares}. 
In particular, we consider the potential effects of Pop~III stars on the timing, shape, and depth of the 21-cm absorption trough. We 
ask when UV emission from the stars themselves can naturally trigger the absorption trough, how rapidly X-ray emission from accreting Pop~III remnant black holes can heat the gas and transform the absorption into emission, and finally whether radio emission from these black holes  can raise the radio background temperature far enough to affect the amplitude of the absorption. 
We also examine the differences between various Pop~III models and how the environments around their birth halos may alter their effect on this signal.

In Section~\ref{sec:popIII_model} we outline the details of our Pop~III semi-analytic model \citep[explained in more detail in][]{mebane_2018}. We describe the details of the 
global cosmic 21-cm signal in Section~\ref{sec:21cm}, and we examine the emission properties of Pop~III remnants in Section~\ref{sec:bh}. We present the results of our model, including 
our best fits to the EDGES signal, in Section~\ref{sec:results}, and we conclude in Section~\ref{sec:conclusion}.

In this work, we use a flat, $\Lambda$CDM cosmology with $\Omega_\text{m} = 0.28$, $\Omega_\text{b} = 0.046$, $\Omega_\Lambda = 0.72$, $\sigma_8 = 0.82$, $n_\text{s} = 0.95$, and $h=0.7$, consistent with the results from \citet{planck}. 

\section{Semi-Analytic Model}
\label{sec:popIII_model}

In this section, we briefly summarize the details of our semi-analytic model for Pop~III star 
formation, which is detailed further in \citet{mebane_2018}. 
We initialize a set of halos in which stars will eventually form at $z=50$ and evolved until $z=6$. Halo masses are chosen to span the range of $10^6 M_\odot$ to $10^{13} M_\odot$ at $z=6$.
The growth rate of these halos is calculated through abundance matching, where 
we assume that halos maintain their comoving number density throughout cosmic time. To do this calculation, we use mass functions found from fits to high 
redshift simulations of dark matter halo growth by \citet{trac_2015}. We note that these assumptions only track the average growth of 
halos, and thus we do not include the effects of mergers. This is consistent with \citet{behroozi_2015}, who find the majority of 
halo growth at these redshifts comes from smooth accretion of material from the intergalactic medium (IGM).

\subsection{Pop~III Star Formation}

Once we have a collection of dark matter halos and their mass histories, we then begin to model their star formation. Because the 
first star-forming halos were very small and consisted only of pristine gas, Pop~III stars were likely very massive and formed in isolation. 
As they formed and subsequently died, they released metals into their birth halos that would eventually allow for the formation of more 
traditional Pop~II stars. Flexibly modeling this transition is a main goal of this model, as these Pop~II halos will be much 
easier to observe than their Pop~III counterparts.

The first step  
in our semi-analytic Pop~III star formation model is to determine the minimum halo mass at which molecular hydrogen 
cooling becomes efficient enough to allow gas clouds to collapse and form stars. Before the first stars form, this is determined by the amount of H$_2$ 
that a halo can form. \citet{tegmark_97} found the required fraction of H$_2$ in a halo at high redshift for cooling to be efficient is given by
\begin{multline}
f_{\text{crit, H}_2} \approx 1.6 \times 10^{-4} \left( \frac{1+z}{20}\right)^{-3/2}
\left( 1 + \frac{10 T_3^{7/2}}{60 + T_3^4}\right)^{-1} 
\\ \times \exp{\left(\frac{512 \text{K}}{T}\right)},
\label{eq:fh2}
\end{multline}
where $T$ is the virial temperature of the halo and $T_3 = T / 10^3$K.  In these halos, molecular hydrogen is formed primarily through the process
\begin{align}
\text{H} + \text{e}^- \rightarrow \text{H}^- + h\nu
\\
\text{H}^- + \text{H} \rightarrow \text{H}_2 + \text{e}^-,
\end{align}
where free electrons catalyze the reaction. At these redshifts, H$^-$ can be destroyed by cosmic microwave background (CMB) photons, 
so a halo's H$_2$ abundance depends on the balance of the formation and destruction rates of the intermediate ion. \citet{tegmark_97} showed the fractional abundance to be a function of the halo's virial temperature (and hence mass)
\begin{equation}
f_{\text{H}_2} \approx 3.5 \times 10^{-4} \text{ }T_3^{1.52}.
\end{equation}
Once a halo's H$_2$ has exceeded the critical fraction shown in eq.~\ref{eq:fh2}, Pop~III star formation can begin.

After the first stars form, however, the minimum halo mass 
to form Pop~III stars is instead determined by the Lyman-Werner (LW) background. The LW band consists of photons 
in the energy range of 11.5 to 13.6 eV, which can photodissociate H$_2$ through the Solomon process \citep[see][]{solomon}. Once star formation begins, 
a large enough LW background is quickly built up to limit Pop~III star formation to only halos massive enough to shield themselves from this background. 
\citet{visbal_2014} find this critical mass to be
\begin{equation}
M_{\text{min}} = 2.5 \times 10^5 \ M_\odot \ \left( \frac{1+z}{26} \right)^{-1.5} 
\left( 1 + 6.96 \left( 4\pi J_{\text{LW}} \right)^{0.47}\right).
\label{eq:mcrit}
\end{equation}
We calculate the specific intensity of the LW background self-consistently from star formation in our model, and we require halos to exceed this critical mass before they can 
begin forming Pop~III stars. We note that Pop~III stars are expected to emit approximately an order of magnitude more total LW photons 
per baryon than Pop~II stars \citep[see][]{schaerer_2002, barkana_2005}. 
Due to the increased UV photon yield, we might expect 
Pop~III stars to have an early, noticeable effect on the cosmological 21-cm background 
through the Wouthuysen-Field effect (see section~\ref{sec:21cm}).

We note that there have been other studies 
that find the minimum mass required for Pop~III star formation to instead be set by the relative streaming velocity between 
baryons and dark matter \citep[e.g.,][]{schauer_2019}. In general, we find that the minimum mass calculated from the Lyman-Werner background is always higher than the 
minimum mass from this streaming effect in our models so we do not include 
the effects of velocity offsets in our study. 
In comparison to \citet{schauer_2019}, our models (including the LW background) have minimum masses comparable to their ``v3" simulation, which corresponds to offset velocities three standard deviations from the mean. Thus only in very rare volumes, or in the very earliest epoch of star formation (well before the absorption trough becomes significant) do streaming velocities significantly affect the minimum mass.

Once we determine the mass range of halos that can form Pop~III stars, we then make the assumption that Pop~III stars form in isolation in these halos due to their feedback and begin 
to add them to the halo. The mass of each star is individually drawn from the chosen IMF, and a new star can only form after the old star dies. If this star 
ends in a supernova that blows out most of the gas, we must wait for the halo to accrete enough new material for star formation to begin again. This typically causes a delay of a few million years between star formation episodes in models 
where supernovae are common.

We allow for a number of different initial mass functions (IMF) for Pop~III stars. Because there have yet to be any observations of this mode of star 
formation, the IMF is largely unconstrained. \citet{mckee_2008} calculate the maximum mass of a Pop~III star by stopping its growth once radiative 
feedback becomes strong enough to limit accretion. They find this maximum mass to scale with the virial temperature of a halo as
\begin{equation}
M_{\text{max}} \approx 145 M_\odot \left( \frac{25}{T_3}\right)^{0.24}.
\label{eq:mmax}
\end{equation}

In our fiducial models we test the case of three separate IMFs. The first, referred to as our ``low'' IMF in the rest of this paper, is a Salpeter-like IMF 
with a minimum mass of 20 M$_\odot$ and a maximum mass given by eq.~\ref{eq:mmax}. Our ``mid'' IMF assumes that, since Pop~III stars grow in isolation, they 
will all be able to reach the maximum mass in eq.~\ref{eq:mmax}. Finally, our ``high'' IMF allows for the possibility of very massive Pop~III stars which form 
from a Salpeter-like IMF spanning the range of 200 M$_\odot$ to 500 M$_\odot$.

Because the halos forming Pop~III stars are so small (we find they can be as small as $10^5 M_\odot$ at the highest redshifts) feedback from Pop~III star 
formation is very important. In particular, supernova feedback plays a vital role in regulating this mode of star formation, as the binding energies of these 
halos are of the order
\citep{first_galaxies}:
\begin{equation}
E_b \approx 2.53 \times 10^{50} \left(\frac{\Omega_\text{m}}{\Omega_\text{m}(z)} \right)^{1/3} 
 \left( \frac{M}{10^6 M_\odot} \right)^{5/3} \left( \frac{1+z}{10}\right) h^{2/3} \text{erg}.
 \label{eq:Eb}
\end{equation}
Since a core-collapse supernova can release kinetic energy on the order of $10^{51}$ erg and a pair-instability supernova can exceed that by an order of magnitude 
(see \citealt{wise_2008} and \citealt{greif_2010}), a single supernova from a Pop~III star can potentially unbind all of the gas in its birth halo (assuming $\sim$1 to 10\% of this kinetic energy
 is able to bind to the gas). 
This has implications for the transition to Pop~II star formation as any metals 
released in a supernova can potentially be carried out of the halo. We use metal yields of Pop~III stars from \citet{heger_2010} and \citet{heger_2002}, which provide 
yields for core-collapse and pair-instability supernovae, respectively. In general, the specific yields of supernovae are not very important, 
as a single supernova provides more than enough carbon and oxygen to 
allow for efficient metal-line cooling assuming at least a few percent of the metals are retained by the halo.

Whenever a supernova occurs in a halo (typically $\sim 5$ Myr after the Pop~III star forms), we assume 10\% of the kinetic energy released binds to the gas in that halo. 
We then track this ejected gas's mass and metallicity, allowing it to reaccrete after a free-fall time. In the intervening time, pristine material from the IGM is still 
allowed to accrete onto the halo. This means that, in our model, multiple generations of Pop~III stars are allowed to form before a halo becomes massive enough to retain the metals 
released in a supernova and transition to Pop~II star formation. We find that halos typically form $\sim 10$ Pop~III stars before becoming 
tightly enough bound to retain enough 
metals and transition. This typically occurs around the same time a halo reaches the atomic cooling threshold at a virial temperature of $10^4$ K.

We note that our fiducial semi-analytic model differs slightly from similar models in that we allow allow multiple generations of Pop~III stars to form in a single halo and these stars form in isolation. For example, 
\citet{visbal_2017} and \citet{jaacks_2017} only allow a single generation of Pop~III stars to form in a halo before transitioning to metal enriched star formation. We investigate the 
differences between these two approaches as well as compare a run of our model with only one generation of Pop~III star formation per halo in \citet{mebane_2018}.

\subsection{Pop~II Star Formation}

The formation of Pop~II stars is also an important part of our model, as Pop~II stars very quickly begin to dominate the LW background and set the minimum halo mass 
for Pop~III star formation. We use the feedback-regulated models of \citet{furlanetto_2017} who test models where either energy or momentum is conserved in 
supernova winds. In these models, 
the star formation efficiency, $f_\ast$, is defined as the fraction of accreting material that will turn into stars and is written as
\begin{equation}
f_\ast \approx \frac{1}{1 + \eta \left( M_h, z\right)},
\end{equation}
where $\eta$, defined as $\dot{M}_{\text{ej}} = \eta \dot{M}_\ast$, relates the star formation rate to the rate at which gas is ejected from the halo due to feedback. In the 
energy-regulated case, $\eta$ is determined by balancing the rate at which kinetic energy is released into the halo 
by supernovae with the rate at which the halo gains binding 
energy from accretion. This can be written as
\begin{equation}
\eta_{\text{E}} = 10 \epsilon_\text{k} \omega_{49} \left( \frac{10^{11.5} M_\odot}{M_h}\right)^{2/3} \left( \frac{9}{1 + z}\right),
\end{equation}
where $\epsilon_\text{k}$ is the fraction of a supernova's kinetic energy which is used to drive a wind, and $\omega_{49}$ is the amount of energy released in supernovae per unit mass of 
star formation in units of $10^{49}$ erg M$_\odot$. In our fiducial model, $\omega_{49} = 1$ and $\epsilon_\text{k} = 0.1$.
In the momentum-regulated case we instead conserve momentum in this calculation and write $\eta$ as
\begin{equation}
\eta_\text{p} = \epsilon_\text{p} \pi_\text{fid} \left( \frac{10^{11.5} M_\odot}{M_h}\right)^{1/3} \left( \frac{9}{1 + z}\right)^{1/2}.
\end{equation}
Here, $\epsilon_\text{p}$ defines the fraction of the momentum released in a supernova that is used to drive winds in the halo and is taken to be $0.2$ fiducially. $\pi_\text{fid}$ 
parameterizes the momentum injection rate from stars formed from a given IMF and is of order unity for a Salpeter IMF (see \citet{furlanetto_2017} for a more detailed derivation of these 
quantities). 
These parameter choices match the observed luminosity functions at $z \ga 6$ reasonably well but offer contrasting extrapolations to higher redshifts and smaller halo masses.

In general, the momentum-regulated case allows for more efficient star formation in low mass halos, which yields a higher LW background. This cuts off Pop~III star formation 
at a much earlier time, as the minimum mass rises very quickly.

\begin{figure*}
    \includegraphics[width=2\columnwidth]{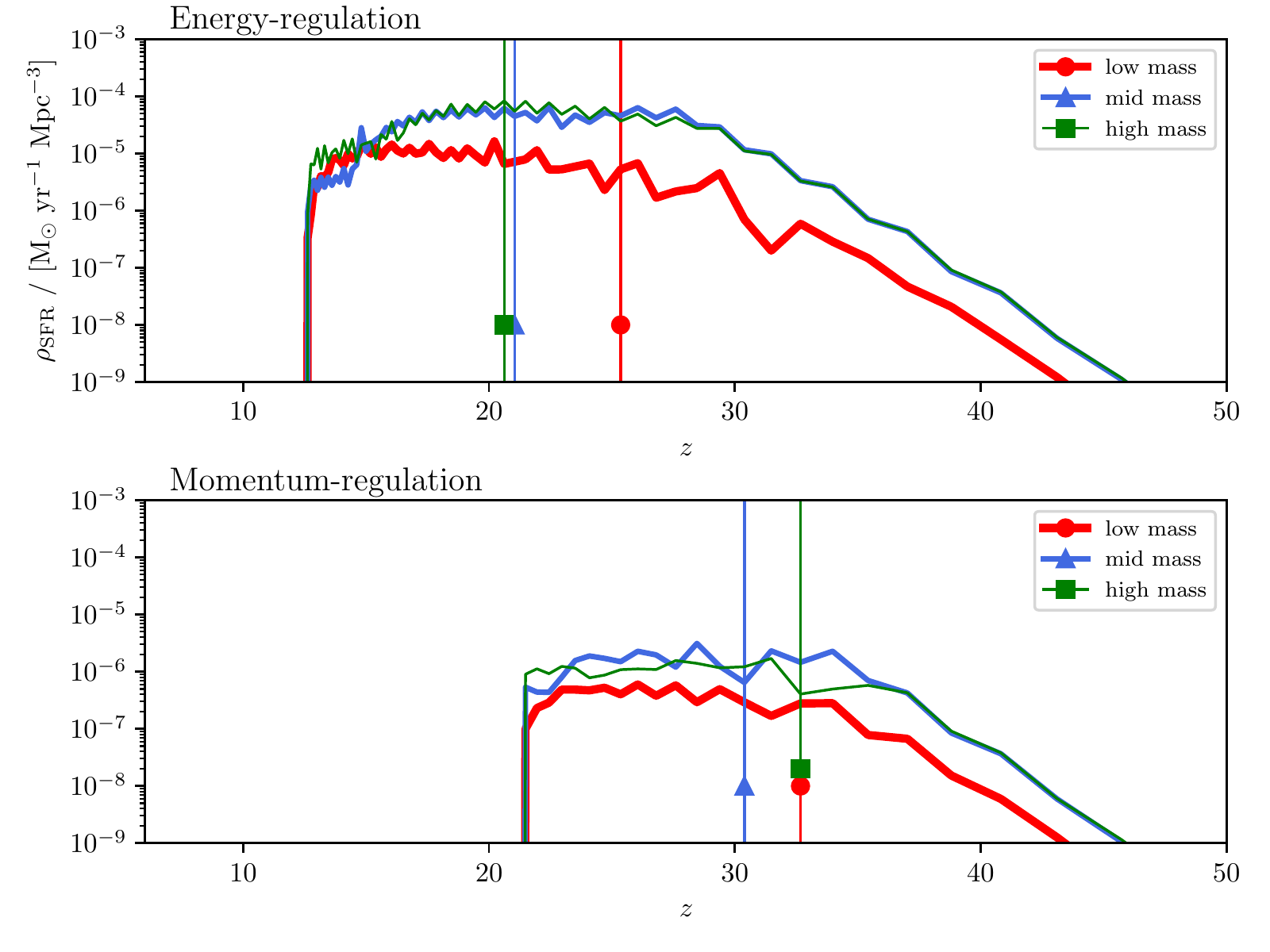}
    \caption{Star formation rate density of minihalo Pop~III stars for a variety of our models. Symbols indicate where Pop~II star formation overtakes Pop~III star formation. The upper panel shows our results for a low mass Pop~III IMF under a variety of different assumptions for the Pop~II and III star formation prescriptions. The bottom panel show a comparison between three different Pop~III IMFs using energy- and momentum-regulated Pop~II star formation, respectively. 
In each case, the minihalo Pop~III era ends when the minimum mass to form stars exceeds the minimum mass for gas to cool via atomic cooling ($\sim 10^4$~K), because we assume the halos transition to Pop~II star formation at that point.
    Note that models which employ momentum-regulated Pop~II star formation will form stars more efficiently in low-mass halos, raising the minimum mass above the atomic cooling threshold faster and transitioning from the minihalo Pop~III phase sooner. We note that we do not include all of these models 
    in the following calculations of the 21-cm global signal, though we show them here to illustrate the range of star formation rate densities our semi-analytic model can produce.}
    \label{fig:sfrd_comp}
\end{figure*}

The results of our semi-analytic model are summarized in Fig.~\ref{fig:sfrd_comp}, which shows the star formation rate density of Pop~III stars for a variety of models.

\section{The Global 21-cm Signal}
\label{sec:21cm}

The 21-cm differential brightness temperature can be written as \citep[][]{furlanetto_2006a}
\begin{eqnarray}
\delta T_b & \simeq & 27 x_\text{HI} \left( 1 + \delta \right) \left( \frac{\Omega_\text{b} h^2}{0.023} \right) \left( \frac{0.15}{\Omega_\text{m} h^2} \frac{1+z}{10} \right)^{1/2} \nonumber \\
& & \left( \frac{T_\text{S} - T_\text{rad}}{T_\text{S}} \right) \text{ mK},
\end{eqnarray}
where $\delta$ is the overdensity, $x_\text{HI}$ is the fraction of neutral hydrogen in the universe, $T_\text{rad}$ is the temperature of the radio background (typically the CMB temperature), 
and $T_\text{S}$ is the spin temperature of neutral hydrogen,
\begin{equation}
T_\text{S}^{-1} = \frac{T_\text{rad}^{-1} + x_\alpha T_\alpha^{-1} + x_\text{c} T_\text{K}^{-1}}{1 + x_\text{c} + x_\alpha}.
\end{equation}
Here, $x_\text{c}$ is the collisional coupling coefficient, $T_\alpha$ characterizes the strength of the Lyman-$\alpha$ background, and $T_\text{K}$ is the kinetic temperature of the gas. $x_\alpha$ is the radiative coupling coefficient quantifying the Wouthuysen-Field effect \citep[][]{wouthuysen, field} and is defined as \citep[][]{chen_2004, hirata_2006}
\begin{equation}
x_\alpha = 1.81 \times 10^{11} (1+z)^{-1} J_\alpha,
\end{equation}
where $J_\alpha$ is the Ly$\alpha$ flux computed from the Pop~III and Pop~II sources in our semi-analytic model. 
We combine the semi-analytic model described above with 
the global 21-cm simulation code \textsc{ares},  
which computes the relevant radiation backgrounds and temperatures and is described in more detail in \citet{ares}, to compute the effects of Pop~III star formation on the global 21-cm 
background.

Recently \citet{edges} reported the first detection of a cosmological 21-cm global signal with the EDGES experiment. 
While this detection has yet to be confirmed by another experiment, it 
shows some features inconsistent with previous theoretical predictions. Specifically, it 
has an absorption feature centered around a frequency of $\sim 78$ MHz or $z \sim 18$. 
This is  earlier than the predictions of most models which study star formation in \emph{atomic} cooling halos \citep[i.e.,][]{mirocha_2016}, suggesting 
that an unknown source of star formation generated a Ly-$\alpha$ background at these early times. 
Also, the absorption feature appears 
to be much deeper than predicted at $\sim$500 mK, potentially implying that the gas had somehow cooled faster than expected for adiabatic expansion
or the existence of another radio background (over and above the CMB) against which the 21-cm absorption occurs. We note that, due to the challenging nature of these observations, there are 
still concerns about the cosmic origin of this signal due to systematics and foreground contamination \citep[e.g.,][]{hills_2018, draine_2018, spinelli_2019, bradley_2019, sims_2019}.

The global 21-cm signal is determined by the factors $x_{\rm HI}$ -- likely near unity throughout the era of Pop~III star formation -- and the temperature.
The latter depends on: (1) the ultraviolet background (through $x_\alpha$), and hence the star formation rate; (2)  the X-ray background, which likely determines $T_K$ because X-rays can propagate large distances through the IGM; and (3) the radio background, if $T_{\rm rad}$ differs from the CMB. The first follows directly from the Pop~III semi-analytic model described in \S \ref{sec:popIII_model}, but the other factors require us to follow the growth of black holes in the early Universe. In the next section, we describe how we supplement our semi-analytic model to do that.

\section{Emission From Black Holes}
\label{sec:bh}

In our semi-analytic model, we self-consistently model the production and growth of black holes from Pop~III stars as well as their X-ray and radio emission. Once the first Pop~III star of the 
appropriate mass in a halo reaches the end of its life, we form a black hole that grows throughout the rest of the run of the model. 
This is  where our choice of IMF has the most effect, as it is thought that a star with mass between $140 \, M_\odot$ and $260 \, M_\odot$ will end its life in 
a pair-instability supernova which will not leave behind a remnant. Similarly, a star with mass in the range $40 \, M_\odot$ and $140 \, M_\odot$ and above $260 \, M_\odot$ will not result in a supernova but will instead collapse directly into a black hole \citep[i.e.,][]{heger_2002, heger_2010}. 

The growth of a single black hole in a halo is governed by
\begin{equation}
\dot{M}_\text{BH} = f_\text{Edd} f_\text{gas} \dot{M}_\text{Edd} + \dot{M}_\text{BH, new},
\end{equation}
where $\dot{M}_\text{Edd}$ is the rate of Eddington-limited accretion, $\dot{M}_\text{BH, new}$ is the rate at which new black holes are produced in the halo, and $f_\text{Edd}$ is the fraction 
of the Eddington limit at which black holes are allowed to accrete (fiducially taken to be 0.1). 
Since we track the amount of gas that is available inside a halo due to accretion and feedback, we do not allow black holes to accrete if there is not sufficient gas in the halo. The factor $f_\text{gas}$ is hence unity if the halo has a gas reservoir available and zero otherwise. Halos typically contain more than enough gas for accretion onto the central black hole, and it is only in the first few million years after a supernova that $f_\text{gas} = 0$.
Since halos in our fiducial model can form multiple generations of Pop~III stars, we make the simple assumption that all new black holes will eventually merge with the central black hole in the halo, so anytime a new black hole is created its mass is added to that of the main black hole. Once a halo transitions to Pop~II star formation, all future growth is from accretion rather than the production of new black holes. 
Note that we do ignore any new black holes that Pop~II galaxies form; in the models we examine, the IGM is already heated significantly by this point, so such black holes do not affect our conclusions.

\begin{figure}
    \includegraphics[width=\columnwidth]{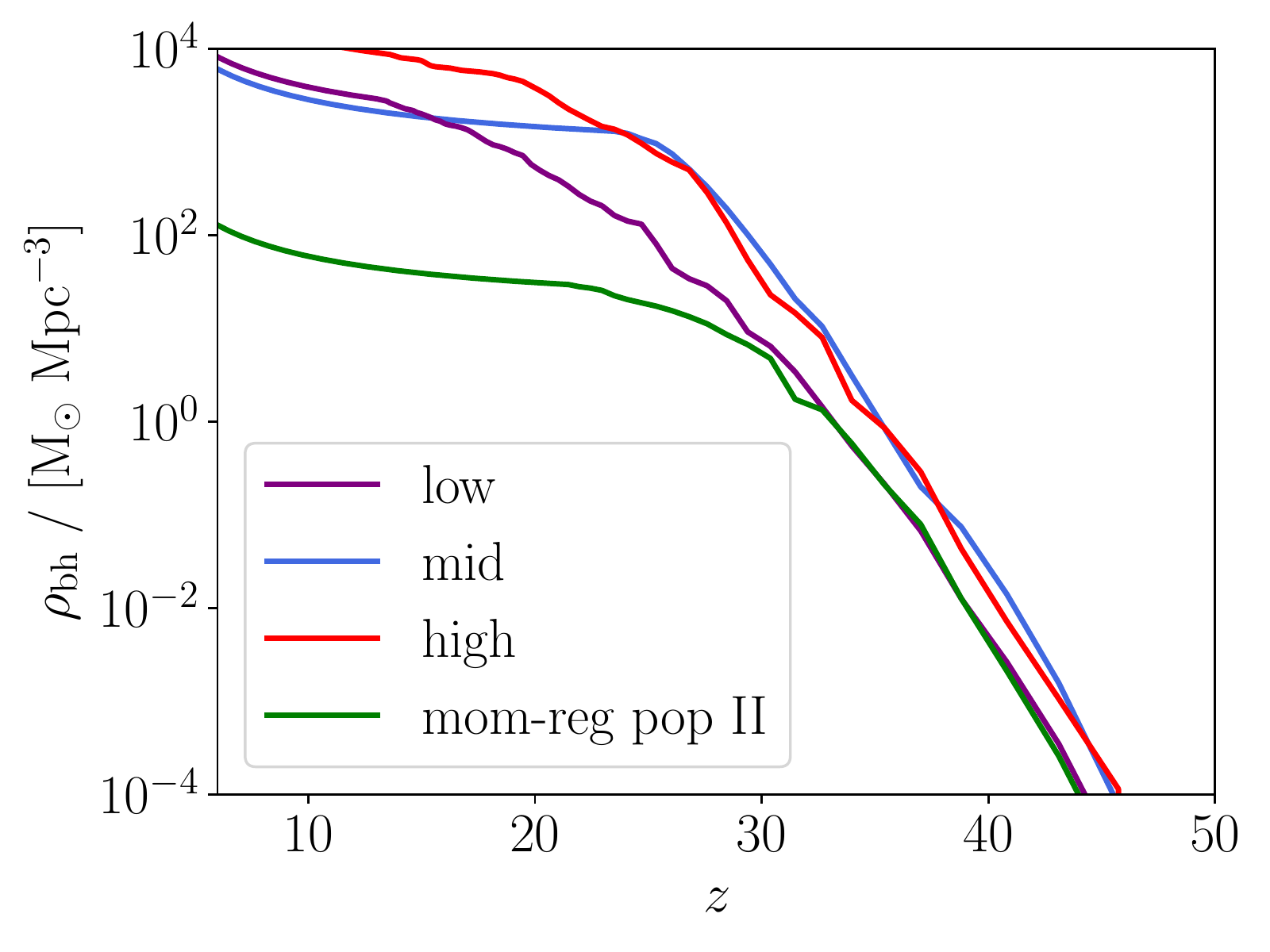}
    \caption{Black hole mass density with an accretion efficiency of 10\%. The break at z$\sim 25$ in the ``mid'' model is due to the slight redshift dependence on the maximum 
    Pop~III mass from \citet{mckee_2008}. At this point, Pop~III stars in new halos are massive enough to end in pair instability supernovae, so they do not leave behind a black hole, 
    and any growth in the black hole mass density is due to accretion. Note that these densities are, at their maximum, at least an order of magnitude lower the mass density of black holes 
    at lower redshifts which have been found to be $\sim 10^5$ M$_\odot$ Mpc$^{-3}$\citep{aller_2002}.}
    \label{fig:bhmd}
\end{figure}

We show the resulting comoving black hole mass density, under several different assumptions in our model, in Fig.~\ref{fig:bhmd}. In all cases, we assume that black holes accrete at 10\% of the Eddington limit. In general, creation of new black holes is dominant over growth via accretion, especially during the early stages of growth when black holes are still small. We note that the specific parameters governing black hole growth in the early universe are not well known (i.e., our assumptions of $f_\text{Edd} = 0.1$ and that all Pop~III black holes in a halo will eventually merge), although our results are at least qualitatively the same for a large range of black hole growth assumptions.

If there is a significant population of accreting Pop~III remnants at high redshift, they should emit in the X-ray as well as the radio. Both such backgrounds have strong implications for the 21-cm signal: an X-ray background will serve to heat the neutral hydrogen in the IGM, decreasing the depth of the absorption trough, but a radio background can increase the depth by raising the background radio temperature \citep{feng_2018}. Moreover, the onset of both backgrounds -- which grow roughly exponentially in the early phases, as shown in Fig.~\ref{fig:bhmd} -- will affect the timing of any 21-cm features. In the next two subsections we explore our models for these backgrounds. 

\subsection{X-ray Heating}
\label{sec:xray}

To study this effect, we self-consistently compute the X-ray background in \textsc{ares} from the Pop~III remnant black hole mass densities calculated in our semi-analytic model. In our model, we assume that individual accreting black holes emit X-rays with a multi-color disk spectrum \citep{mitsuda_1984} with luminosity
\begin{equation}
L_\text{BH} = 1.26 \times 10^{38} \, f_X \ \text{erg s}^{-1} \left( \frac{M_\text{BH}}{10 \text{M}_\odot}\right) \left( \frac{f_\text{Edd}}{0.1}\right) \left( \frac{f_{0.5-8}}{0.84}\right),
\end{equation}
where $f_{0.5-8}$ is the fraction of the energy emitted in the 0.5 keV to 8 keV band for a 10 M$_\odot$ black hole. Here $f_X$ is an arbitrary scaling factor that allows us to vary the overall efficiency of X-ray production during accretion.

We note that a significant portion of the X-ray background could be suppressed if halos contain a large enough column density of neutral hydrogen and helium. \citet{aew_2018} find that a column density of 
$\sim 10^{23}$ cm$^{-2}$ is required to block X-ray emission from a halo. We find that halos in our semi-analytic models can only reach these column densities if they contain close to 
the cosmic baryon fraction of baryons and if all of these baryons can quickly cool onto the center of the halo. While gas can cool fairly easily onto the halo, it is difficult for 
haloes to keep enough gas in the halo without it being disrupted by Pop~III supernovae. 
Nevertheless, we explore the effects of local absorption more thoroughly in section~\ref{sec:results_depth}.

\subsection{Excess Radio Background}
\label{sec:radio}

We also include a self-consistent calculation of the radio background produced by the accreting black holes  in our semi-analytic model using the \textsc{ares} code. The brightness temperature of this background at $\nu = 1420.41$ MHz for a given density of black holes is
\begin{equation}
T_\text{rad} (z) = \frac{c^2 J_\nu(z, f_R)}{2 \nu^2 k_B},
\end{equation}
where $J_\nu(z, f_R)$ is the specific intensity of the radio background experienced by clouds of neutral hydrogen at redshift $z$. The specific mechanism for the production of this radio background from accreting block holes is not well understood, so we follow \citet{aew_2018} who use empirical trends to 
calibrate this background (see, for example, \citealt{merloni_2003} for a discussion of a ``fundamental plane'' of black hole X-ray and radio emission). They calculate $J_\nu(z, f_R)$ as
\begin{equation}
J_\nu(z, f_R) = \frac{c}{4 \pi} (1+z)^3 \int_z^\infty \epsilon \left( \nu \frac{1 + z'}{1 + z}, z', f_R \right) \frac{dz'}{(1+z') H(z')},
\end{equation}
where $\nu = 1420.41$ MHz and $\epsilon$ is the emissivity of the radio background,
\begin{multline}
\epsilon \left(\nu, z, f_R \right) \approx 1.2 \times 10^{22} \left( \frac{f_R}{1} \right) \left( \frac{\rho_\text{bh}}{10^4 h^2 M_\odot \text{Mpc}^{-3}} \right) \\ \times \left( \frac{\nu}{1.4 \text{ GHz}} \right)^{-0.6} \text{ W Hz}^{-1}h^3 \text{Mpc}^{-3}.
\label{eq:radio_emis}
\end{multline}
Here $\rho_\text{bh}$ is the black hole mass density computed from our semi-analytic model (see 
Fig.~\ref{fig:bhmd}).\footnote{Note that we have simplified eq.~\ref{eq:radio_emis} relative to the treatment of \citet{aew_2018}, as details such as the duty cycle of accretion, $f_\text{duty}$, are degenerate with our semi-analytic model (which natively tracks when gas is available in a halo for accretion). Similarly, we have omitted factors relating to the fraction of radio loud accreting black holes and absorbed this into $f_R$.} We also introduce a new 
dimensionless parameter, $f_R$, which boosts the emissivity of radio emission from accreting black holes
relative to the empirical calibration. 
We vary this normalization parameter to fix the excess radio background generated from the black holes.
We note \citet{mirocha_2019} 
require $f_R \sim 100$ to match the 
amplitude of the EDGES signal, although they do not include the effects of Pop~III star formation.  

\begin{figure}
    \includegraphics[width=\columnwidth]{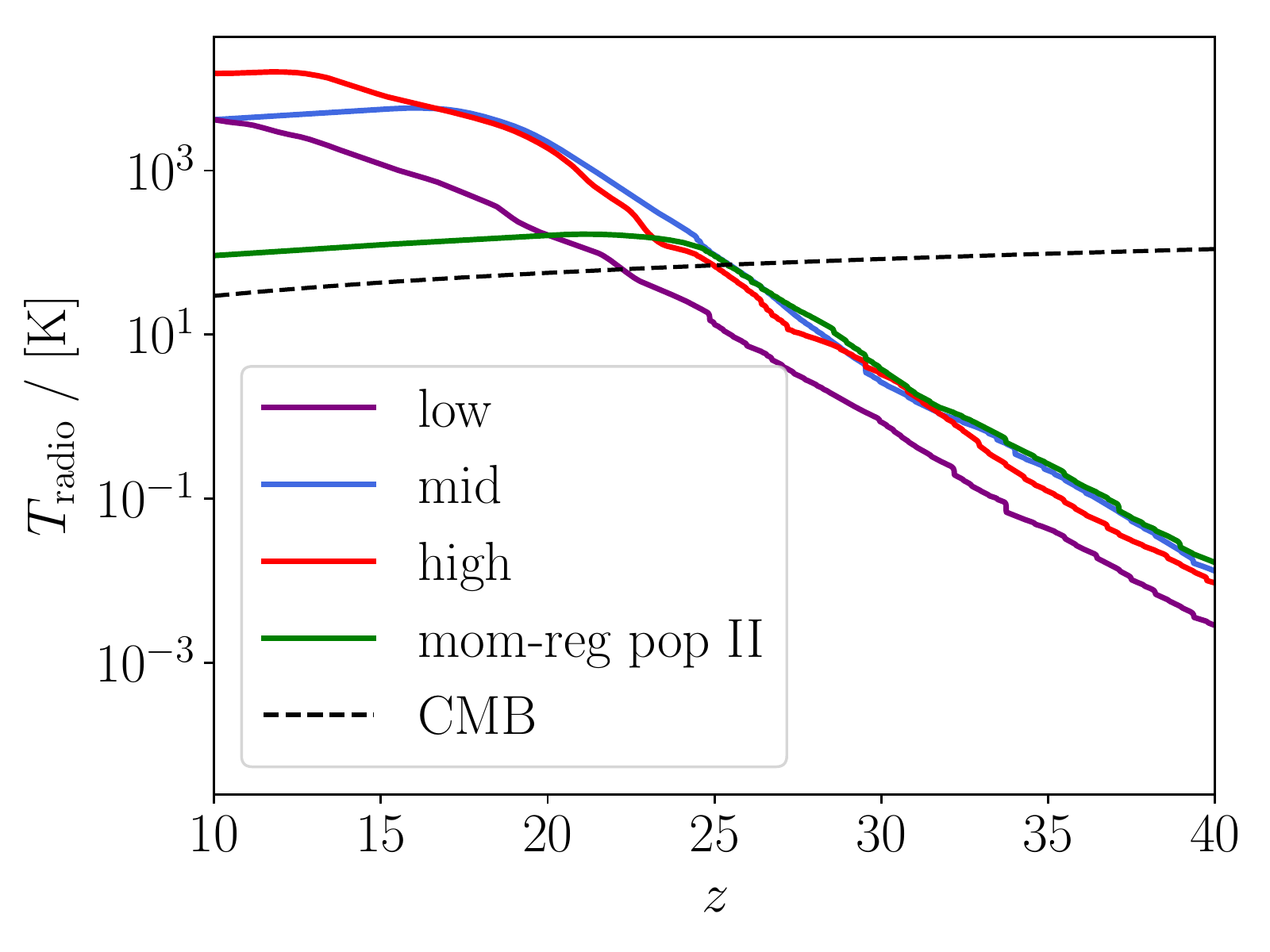}
    \caption{Temperature of the excess radio background computed from the black hole mass densities in our semi analytic model. For this calculation we have set $f_R = 50$. The dashed black curve shows the temperature of the CMB. 
    All models assume energy-regulated Pop~II star formation besides the green curve, which assumes momentum-regulation. If Pop~III remnant black holes are able to efficiently accrete and emit a radio background 
    this strongly, the resulting brightness temperature of the dominant 
    radio background can be much higher than that of the CMB.}
    \label{fig:tbg}
\end{figure}

Fig.~\ref{fig:tbg} shows the brightness temperature of this radio background for a number of  
scenarios generated using our model with $f_R = 50$.\footnote{We show results for the apparently extreme choice of $f_R=50$ because we will find later that this is roughly the value required to reproduce the depth of the EDGES signal.}
Assuming accreting Pop~III remnant black holes are able to emit in the radio in this way, we find a temperature that can be as much as $\sim 100$ times the CMB temperature at the relevant redshifts. Since the depth of the primary absorption feature in the global 21-cm signal 
measures the difference between the spin temperature of the neutral hydrogen gas and the brightness temperature of the dominant radio background, a background such as this would greatly 
enhance the strength of such a feature.

Models which end Pop~III star formation earlier, such as our momentum-regulated Pop~II models, tend to have a much lower radio background temperature that eventually begins to fall like the 
temperature of the CMB due to redshifting. This is due to our assumption that all Pop~III black holes will eventually merge combined with the mass dependence of Eddington-limited accretion. Since halos spend a significantly shorter time in the Pop~III phase in these models, their black holes do not reach a high mass early on due to mergers. This causes them to grow very slowly through smooth 
accretion, as they do not get the same ``bump'' in mass that Pop~III black holes would get if their halos formed many more generation of Pop~III stars (as in our energy-regulated models).

\section{Results}
\label{sec:results}

We are now ready to examine our Pop~III model's implications for the global 21-cm background. We will consider two important aspects of the signal: its \emph{timing}, which includes both the point at which the absorption trough reaches its minimum but also the shape of the signal around it, and its \emph{depth}. The former is important for any model of the early Universe, because it has direct implications for measuring the earliest generations of star and black hole formation. The depth is particular interesting in the context of the EDGES measurement: specifically, we are interested in whether Pop~III models can provide a self-consistent explanation for the anomalous depth of the EDGES absorption trough.

Even within the context of our semi-analytic model, the uncertainties in Pop~III star formation are large enough that we cannot examine every possible scenario in detail. We therefore focus on two qualitative questions, rather than attempting to provide quantitative constraints from the EDGES claim.  
First, is it 'natural' for PopIII models to produce an early absorption trough, given that PopII-only models do not predict such early features (\citealt{mirocha_2019})?
That is, do a broad range of parameters and assumptions produce such a trough? Second, is it \emph{possible} for ``fine-tuning" to produce surprising results -- in particular, the large amplitude of the EDGES claimed detection?

\subsection{The Timing of the Global 21-cm Signal}
\label{sec:results_timing}

In this section we investigate the timing of the absorption feature of the global 21-cm signal. In general, we expect the addition of Pop~III star formation to push the signal to higher 
redshifts. If early Pop~III star formation is able to build up a high enough Ly$\alpha$ background, we should expect Wouthuysen-Field coupling to occur earlier and drive the spin temperature 
of the gas toward the kinetic temperature at earlier times. This will cause the signal to move into absorption earlier, potentially to $z \sim 18$ where the EDGES signal reaches its minimum.

Figures~\ref{fig:radio} and \ref{fig:xray} show examples of the global signal computed from the energy-regulated models with a low-mass Pop~III IMF described in section~\ref{sec:popIII_model} 
with the EDGES signal overplotted. 
In these figures, we primarily vary the parameters controlling the X-ray and radio backgrounds, $f_R$ and $f_X$. We hold all other parameters constant for simplicity, taking $f_\text{Edd} = 0.1$, 
assuming no absorption of X-rays due to a local column density of neutral gas, and our ``low'' mass Pop~III IMF. 
We do vary the Pop~II prescriptions: the solid lines assume energy regulation, while the dot-dashed curves assume momentum regulation.
We note that our models with energy-regulated Pop~II star formation provide the best fits to the EDGES signal, as the momentum-regulated models do not have a 
large effect on the global signal unless their Pop~III emission is tuned to be very high. This is because the star formation efficiency in low mass Pop~II halos in the momentum-regulated case is 
much higher, causing a high LW background which shuts off Pop~III star formation sooner.

Regardless of the depth of the signal, we find that, 
over a broad range of parameter space and model assumptions, an era of Pop~III star formation can generate a large enough UV background to trigger Wouthuysen-Field coupling and hence a 21-cm  absorption trough. 
In this context, the timing of the EDGES trough \emph{does} occur naturally for many Pop~III star formation scenarios, in contrast to scenarios that assume star formation in $z > 15$ galaxies follows the same trends as in populations at much lower redshifts \citep{mirocha_2019}. As a result, if increasingly efficient Pop~II star formation can be ruled out by future galaxy surveys, the global 21-cm signal reported by EDGES would provide strong evidence of Pop~III star formation in the early Universe. 
We do note, however, that regardless of our parameter choices our fiducial model does not match the 
\emph{shape} of the signal. 
Specifically, the EDGES team found an absorption trough with a flat bottom and a sharp rise out of absorption. Our models have sharp declines at the onset of absorption but only gradually recover, as also found by \citet{mirocha_2019}. This gradual rise is a natural consequence of the pace of structure formation 
and the physics of the 21-cm line. Structure formation proceeds exponentially at these times, but the Wouthuysen-Field effect is essentially a threshold process, so that the spin temperature approaches the kinetic temperature very rapidly once $x_\alpha > 1$. However, X-ray heating is a continuous process, so that the recovery is significantly slower (see also \citealt{mirocha_2017}). Thus, if the EDGES result is confirmed, it will require rapid time evolution in the parameters of the luminous sources. Section \ref{sec:results_param} describes one way in which this might happen -- in that scenario, features in the stellar mass-remnant black hole mass relation induce a rapid change in black hole formation.

\subsection{The Depth of the Global 21-cm Signal}
\label{sec:results_depth}

In addition to the timing, we are also concerned with the depth of the absorption trough, particularly because of the extreme amplitude measured by EDGES. One way to 
increase the amplitude is to allow for gas in the IGM to cool faster than expected for an adiabatically expanding universe. A possible mechanism for this to occur is if hydrogen atoms can interact with dark matter to some degree \citep[e.g.,][]{barkana_2018}. Regardless of the specific mechanism,  
one can fine tune this phenomenon to achieve the required depth of the EDGES signal after the proper timing has been attained \citep[see, for example, the parameterized approach to cooling in][]{mirocha_2019}. In order to match both the timing and depth of the EDGES signal \emph{self-consistently} with Pop~III stars, however, we must look at the specific properties of X-ray and radio emission from Pop~III remnants in our model.

\begin{figure}
    \includegraphics[width=\columnwidth]{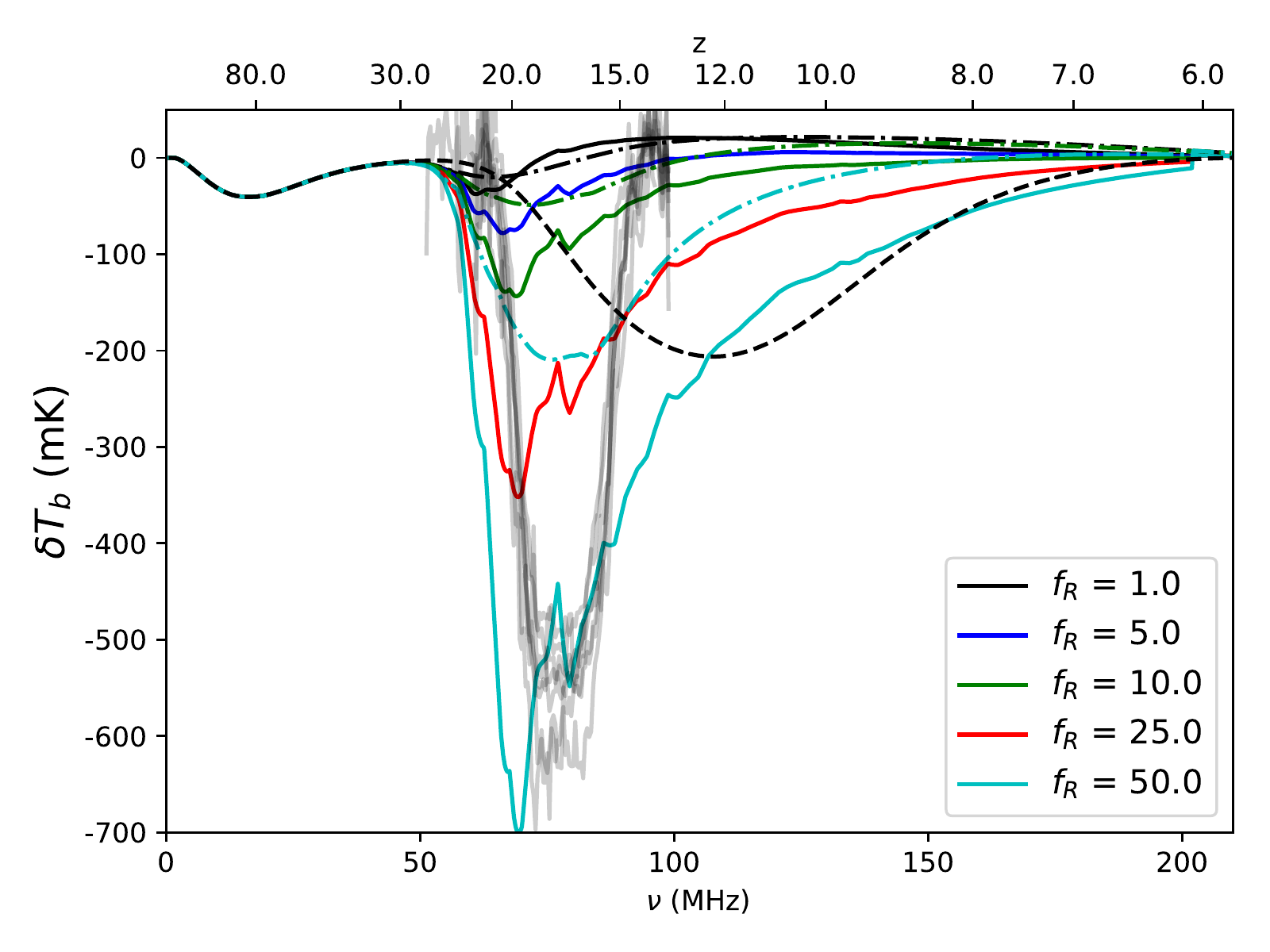}
    \caption{Effects of $f_R$ on the global signal with our low mass Pop~III IMF and energy-regulated Pop~II star formation. The reported EDGES signal fits are overplotted in grey. All models provide a reasonably good fit to the timing of the 
    signal, and the depth of the signal is best fit by boosting the radio emission by a factor of $\sim 50$. In this model, $f_X = 1$, and there is no excess cooling of the IGM. The dashed line shows a model without Pop III for comparison, and the dashed-dotted lines show our results with momentum-regulated Pop~II star formation. In general, it is much harder for us to achieve the large depths 
    of the EDGES signal in these momentun-regulated models.}
    \label{fig:radio}
\end{figure}

As discussed in section~\ref{sec:radio}, accreting Pop~III remnant black holes could produce a high enough radio background to dominate over the CMB and increase the reference radio 
brightness temperature in the global 21-cm signal by factors of up to $\sim 100$. Fig.~\ref{fig:radio} shows the effects of this background for various values of $f_R$ while holding 
the boost to X-ray emission constant at $f_X = 1$. In these models, the best fit to the EDGES signal occurs around $f_R \sim 50$, which reproduces both the strength and timing of the 
signal but does not match the shape.
In particular, note that the 21-cm signal remains in absorption throughout reionization. As shown in Fig.~\ref{fig:tbg}, models with large $f_R$ increase the radio background to $T_R \ga 1000$~K. Seeing the 21-cm signal in emission would require gas heating above this value, which is only possible when $f_X$ is also very large.

\begin{figure}
    \includegraphics[width=\columnwidth]{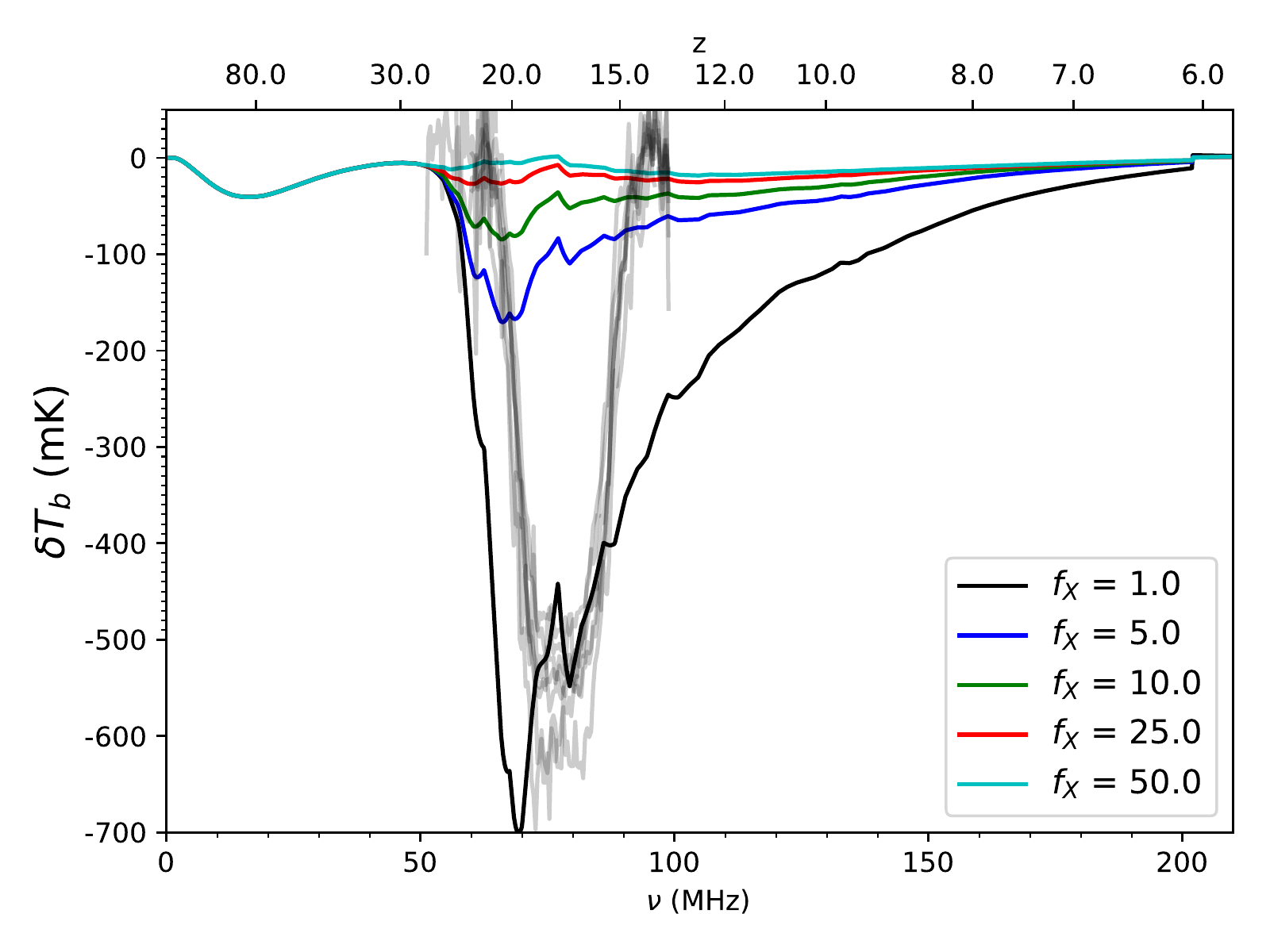}
    \caption{Effects of $f_X$ on the global signal with our low mass Pop~III IMF. Here we have set $f_R = 50$ to better show the effects of an increasingly strong X-ray background. We do not 
    include a neutral hydrogen column density in this case. In general, stronger X-ray backgrounds will heat the IGM, lessening the depth of the absorption signal.}
    \label{fig:xray}
\end{figure}

These same accreting black holes should also produce X-rays, however, which will heat the surrounding gas and weaken the absorption feature of the signal. This is shown 
in Fig.~\ref{fig:xray}, where we have set our radio boost parameter to it's maximal value of $f_R = 50$ to best see the effects of X-ray heating. In general, allowing X-ray emission at high 
levels can significantly lessen the strength of the signal, causing the absorption feature to completely disappear in the most extreme cases. If Pop~III star formation were found to be the 
cause of the EDGES signal, we would need a way to allow accreting Pop~III remnant black holes to emit efficiently in the radio but not in the X-ray.

\begin{figure}
    \includegraphics[width=\columnwidth]{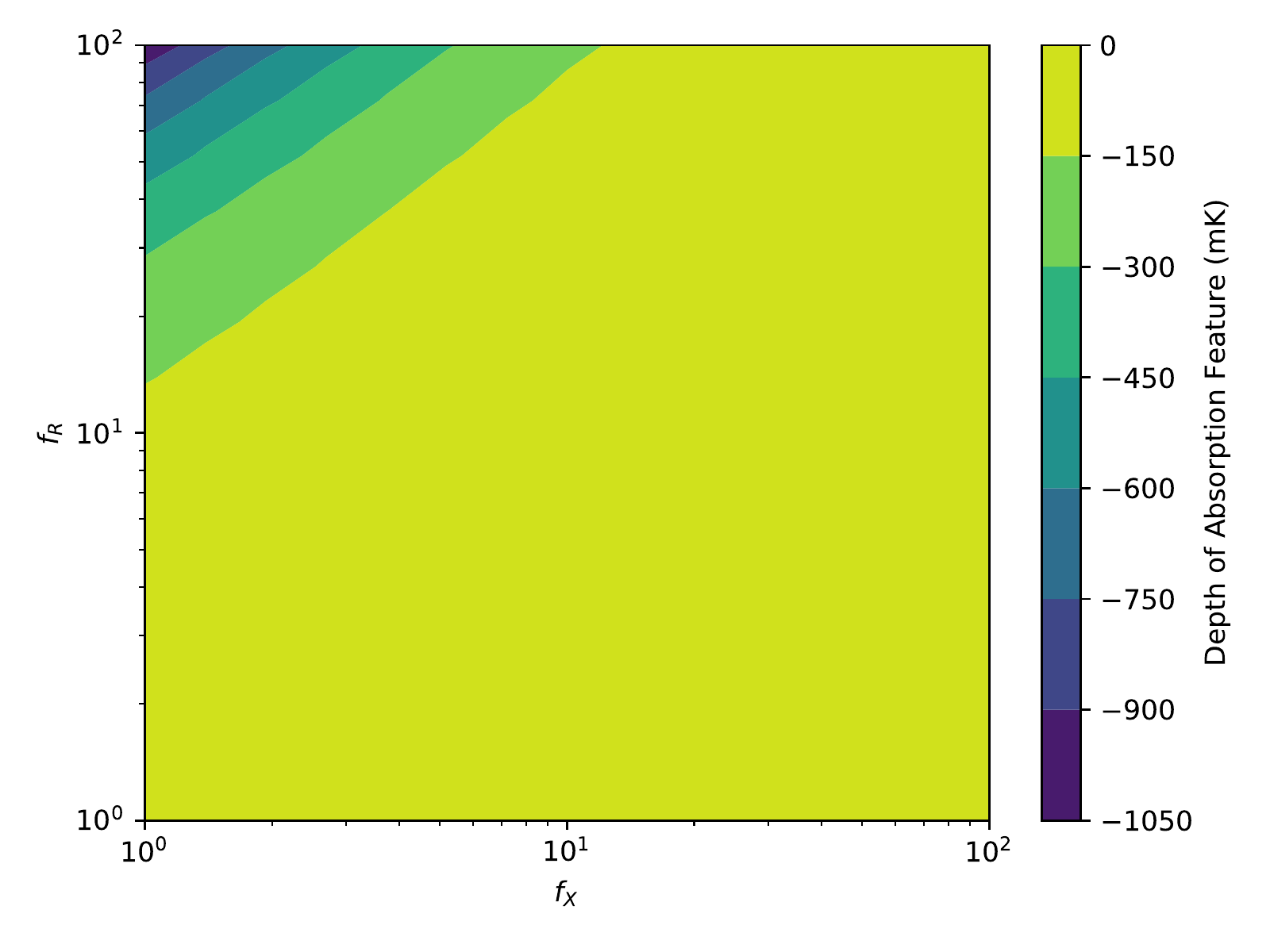}
    \caption{Grid of models with varying values of $f_X$ and $f_R$. Depths similar to those of the edges signal ($\sim -500$ mK) can only occur in our models when $f_R$ is high and $f_X$ is low.}
    \label{fig:grid_IMF1}
\end{figure}

This idea is shown in Fig.~\ref{fig:grid_IMF1}, which 
displays the peak amplitude of the absorption trough from a grid of 225 models varying $f_R$ and $f_X$ from 1 to 100. In general, we find that we can only reproduce the depth of the EDGES 
signal in cases with high $f_R$ and low $f_X$. As discussed in section~\ref{sec:xray}, this could be possible if halos 
have very strong radio emission but also 
have a high enough column density of neutral hydrogen and helium 
to block X-ray emission from the accreting black holes, although 
such high column densities are difficult to obtain in Pop~III halos in our semi-analytic model. 
However, they are easier to imagine in larger halos that have transitioned to Pop~II star formation, and these halos would still contain the Pop~III remnant black holes formed previously.

\begin{figure}
    \includegraphics[width=\columnwidth]{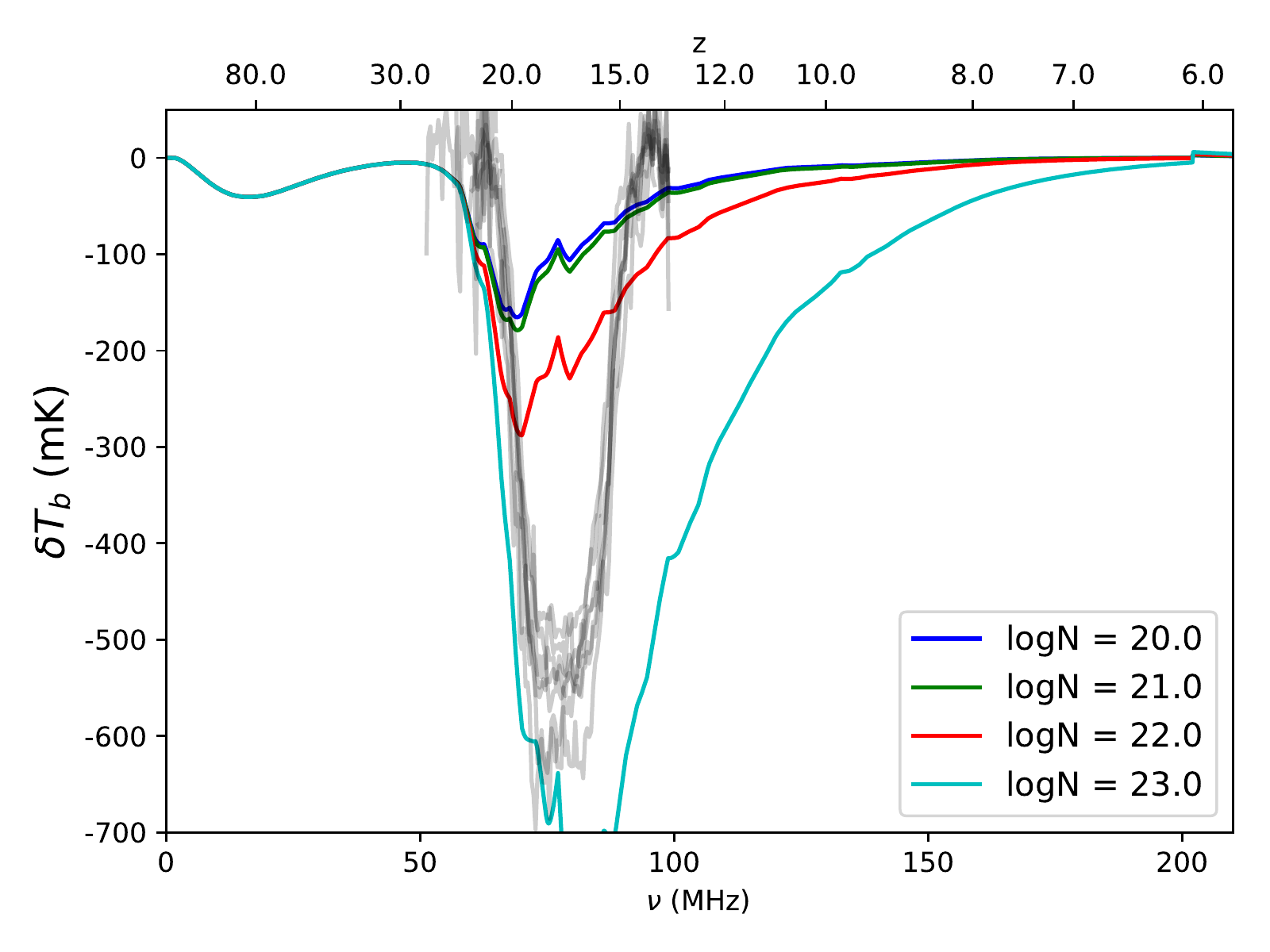}
    \caption{Column density effects on the global signal. At lower column densities, low energy X-rays can easily escape the halo and efficiently heat the surrounding IGM. As the column density 
    gets higher, however, these X-rays are blocked and X-ray heating of the IGM becomes inefficient. This causes a significantly deeper absorption feature that takes a much longer time to rise 
    out of its minimum. Here we have set $f_X = 1$ and $f_R = 10$ to better fit the EDGES signal due to the lessened effect of X-ray emission.}
    \label{fig:logN}
\end{figure}

In order to test the effect of column densities of neutral gas on X-ray emission, we apply various column densities to our Pop~III halos in \textsc{ares} ranging from $N = 10^{20}$~cm$^{-2}$ to $N = 10^{23}$~cm$^{-2}$ in Fig.~\ref{fig:logN}. At the lower column densities there is not much of a difference in the global signal, as lower energy X-rays are able to escape the halo and 
heat the IGM effectively. As we progress to higher column densities, only the higher energy X-rays can escape the halo and are unable to efficiently heat the IGM. In these cases we have set $f_R = 10$ in order to better match the depth of the EDGES signal. This is due to the decreased effect of X-ray heating from neutral gas requiring less radio emission to produce a deep signal. Similar to \citet{aew_2018}, we 
find that a column density of $N \sim 10^{23}$~cm$^{-2}$ is required to fully block this emission and stop X-ray heating of the IGM. 
As we have mentioned above, however, such high column densities require that Pop~III haloes retain all of their gas, which is not likely according to most feedback prescriptions.

\subsection{Parameter Dependence}
\label{sec:results_param}

Most of our models shown 
so far used our energy-regulated Pop~II star formation models which have 
less efficient Pop~II star formation than the momentum-regulated models. We find that, in general, 
the momentum regulated models are able to reproduce the timing of the EDGES signal in a similar fashion, but we must increase $f_R$ to much higher values while still keeping $f_X$ low in 
order to reproduce the depth. This is simply due to the decreased mass density of black holes seen in the momentum regulated models (Fig.~\ref{fig:bhmd}), as the Lyman-Werner background is much higher causing 
Pop~III star formation to end earlier (see Fig.~\ref{fig:sfrd_comp}). This effect is shown in Fig.~\ref{fig:radio}, where we have plotted example global 
signals varying $f_R$. 

We note that our assumption of multiple generations of Pop~III star formation in the same halo has large consequences for the effect of this mode of star formation on the 21-cm background. In 
\citet{mebane_2018} we find that only allowing a single Pop~III star to form in a halo will lower the maximum star formation rate density of the Pop~III phase by at least an order of magnitude 
to similar levels as our momentum regulated models. Thus, in these models, it is much more difficult to attain the required depth to match the EDGES signal. 
Similar models which form a large 
mass of Pop~III stars but in a single generation \citep[e.g.,][]{visbal_2017, jaacks_2017} should, in principle, still be able to attain this depth with similar assumptions for X-ray and radio emission as their Pop~III star formation rate densities are comparable to those of our fiducial models.

\begin{figure}
    \includegraphics[width=\columnwidth]{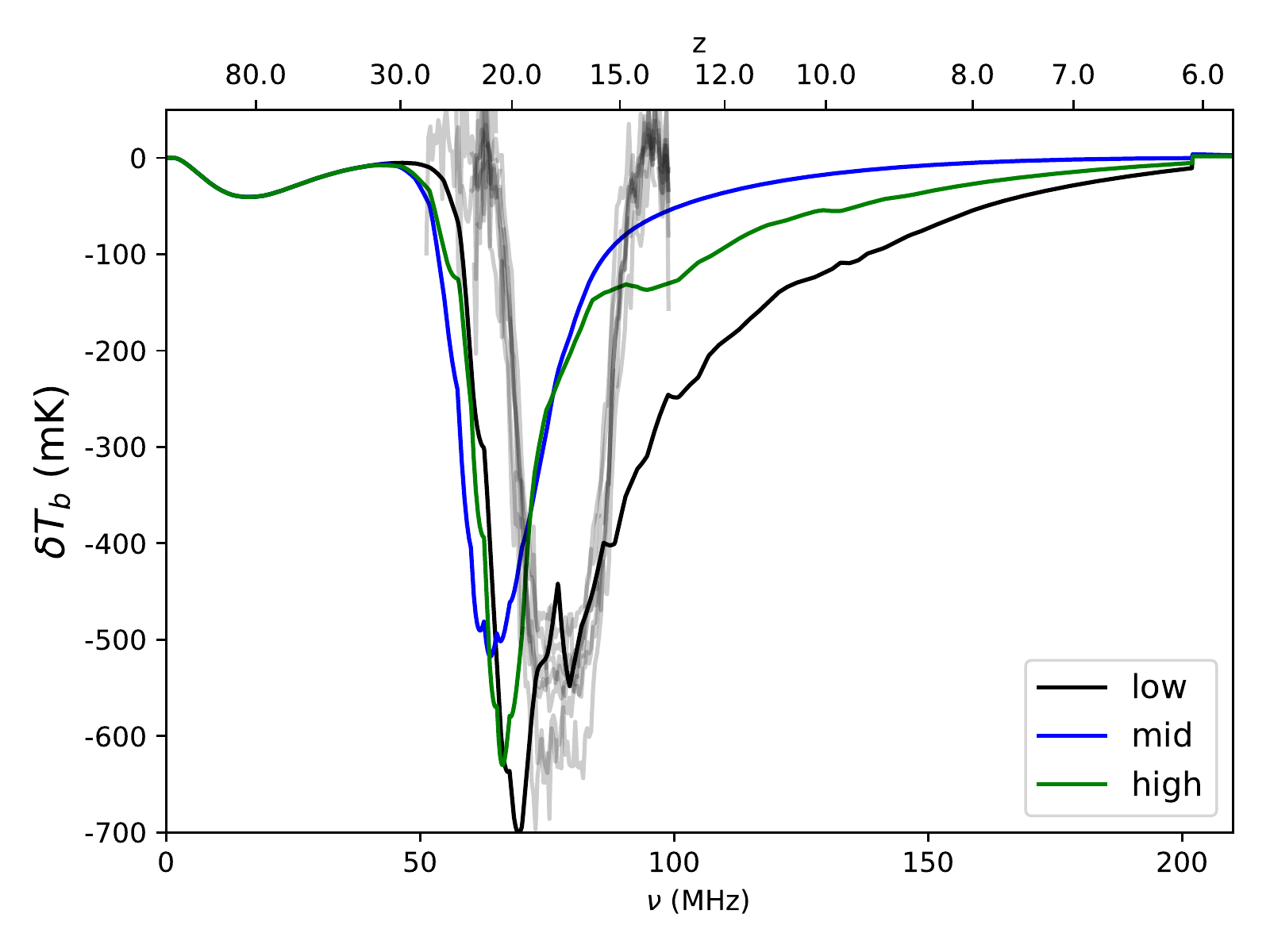}
    \caption{Global signals for varying Pop~III IMFs. We have set $f_X = 1$ in all cases and have tuned $f_R$ such that each model provides a rough fit to the depth of the EDGES signal. For the 
    low mass case, as discussed in section~\ref{sec:results_depth}, $f_R = 50$. In the mid and high mass cases we have set $f_R = 15$.}
    \label{fig:gs_IMFs}
\end{figure}

We also 
consider the effects of our choice of IMF for Pop~III star formation. All previous plots have used our fiducial, low-mass ``low'' IMF, but we show the effects of the ``mid'' and ``high'' 
IMFs (described in section~\ref{sec:popIII_model}) in Fig.~\ref{fig:gs_IMFs}. Generally speaking, the higher mass IMFs which produce larger black hole mass densities cause a stronger absorption 
feature in the global signal. For example, 
matching EDGES with the low mass IMF requires $f_R \sim50$, while the mid and high mass IMFs require only $f_R \sim15$. 
This is simply because larger-mass black holes can grow faster and hence produce more emission.

This is not universally true, however, as IMFs that produce stars in the mass range of $140 M_\odot$ to $260 M_\odot$ are thought to end their lives in a 
pair-instability supernova that will not leave behind a remnant. One potentially interesting difference between our ``mid'' and ``high'' IMF cases is that, while the depth of both signals 
is roughly the same, the ``mid'' IMF case has a sharper rise out of the absorption feature. This is due to the redshift and halo mass dependence of the maximum Pop~III mass calculated by 
\citet{mckee_2008}. This IMF model assumes that \emph{all} Pop~III stars 
form at this maximum mass which, when combined with our self-consistently computed minimum halo mass for Pop~III star formation, 
implies that, after a certain redshift (depending on the model), every new Pop~III star forms in the range for pair instability supernovae. This can be seen in Fig.~\ref{fig:bhmd} where the ``mid'' case 
plateaus after 
$z \sim 30$. Since all new stars at this point will end in a pair instability supernova, any growth in the black hole mass density after this point is from accretion of new material rather than the creation of new black holes. 
As a consequence, Fig.~\ref{fig:tbg} shows that past this point the radio background temperature begins to decline at $z \sim 20$ while the temperature continues to rise in the other scenarios. This decline causes the difference in temperatures between the spin temperature of the gas and the radio background to decrease, allowing for the 
signal to rise more quickly out of absorption. The ``high'' IMF case also has a much quicker rise out of absorption than our fiducial ``low'' case, although to a lesser extent than the ``mid'' model. Again, the reason for this can be seen in Fig.~\ref{fig:tbg} where the radio background temperature for this model has begun to level out. 
While our choice of IMF does not strongly effect the \emph{timing} of the signal, it does indicate that the mass of individual Pop~III stars could have an 
effect on the \emph{shape}.

\section{Conclusions}
\label{sec:conclusion}

We have presented a model to test the effects of Pop~III stars and their remnants on the global 21-cm signal. By combining a semi-analytic model for Pop~III star formation \citep{mebane_2018} with 
a global 21-cm simulation code \citep{ares}, we have found that Pop~III stars can have a large effect on this signal if these stars 
(and their remnants) are allowed to form and emit efficiently. By altering the Pop~II and Pop~III star formation prescription as well as the 
properties of radio and x-ray emission from accreting Pop~III remnants, we are able to drastically alter the shape and depth of the signal.

Specifically, we find:
\begin{enumerate}
\item Our Pop~III models with energy-regulated Pop~II star formation generate a high enough Lyman-$\alpha$ background at the proper time to produce an 
absorption feature at around the right time to be consistent with the EDGES signal. 
\item Allowing for a radio background from Pop~III remnants can significantly increase the strength of the global 21-cm signal. Similarly, an X-ray background from Pop~III remnants can significantly heat the IGM and decrease the strength of the signal. Thus reproducing the depth of the EDGES signal requires Pop III black hole remnants to emit far more efficiently in the radio than in the X-ray, relative to local black holes.
\item Our choice of Pop~III IMF can alter the shape of the signal without strongly affecting the timing or depth. In particular, our ``mid'' IMF model which forms stars from a sharply peaked IMF 
around $140 \, M_\odot$ is able to produce the sharpest signal.
\end{enumerate}

We also compare our results to the recent detection from the EDGES experiment. In general, we find that 
a broad subset of our models for Pop~III star formation are able to match the timing of the EDGES signal quite well, regardless of our choice of IMF and star formation prescription. 
Since the timing of the signal is mostly dependent on star formation at higher redshifts, the most important aspect our models in this regard is the timing of the formation of the very first Pop~III stars. This 
generally happens before feedback specific to the IMF or Pop~II star formation prescription has a large effect, so the most important factors are the physics of molecular hydrogen formation and 
cooling that sets the timing and halo mass range of the first Pop~III stars. Altering other parameters such as the efficiency of accretion on Pop~III black holes, the masses of Pop~III stars, or the number of generations of Pop~III star formation can vastly affect the shape and depth of the signal without altering the timing. Models including both a self-consistent radio and X-ray background can reproduce the depth of the EDGES signal only if $f_R$ is much greater than 
$f_X$. One potential way to explain this is if Pop~III halos have a high enough column density of neutral gas to block much of the X-ray emission 
from accreting black holes. This explanation is hard to 
motivate with our models, however, as we find that Pop~III halos are constantly blowing out much of their gas, 
so that the local column density of neutral hydrogen is likely to be quite small.

While all of the Pop~III models presented here produce a signal that agrees with the timing of the EDGES detection, it is also possible to produce a different timing by altering some of 
the parameters governing this mode of star formation. For example, we could produce a much later trough if we drastically reduce the efficiency of Pop~III star formation. This could be the case if only one generation of Pop~III stars is allowed to form per halo or if the delay between generations of star formation is much larger due to our assumptions regarding accretion after a supernova 
clears the gas out of a halo. The likelihood of these scenarios is discussed in more detail in \citet{mebane_2018}.

\section*{Acknowledgements}

R.H.M would like to thank Andrei Mesinger and Yuxiang Qin for many helpful discussions. J.M. acknowledges support from a CITA National Fellowship. This work was supported by the National Science Foundation through award AST-1812458 and by NASA through award NNX15AK80G. This material is based upon work supported by the National Science Foundation under Grant Nos. 1636646 and 1836019 and institutional support from the HERA collaboration partners.  This research is also funded in part by the Gordon and Betty Moore Foundation. In addition, this work was directly supported by the NASA Solar System Exploration Research Virtual Institute cooperative agreement number 80ARC017M0006. This work relied on the Python packages numpy9 \citep{numpy} and matplotlib10 \citep{matplotlib}.




\bibliographystyle{mnras_edit}
\bibliography{refs} 








\bsp	
\label{lastpage}
\end{document}